\documentclass[letterpaper, 10 pt, conference]{ieeeconf}
\usepackage{amsmath}
\usepackage{graphicx}% Include figure files
\usepackage{epsfig}% Include figure files
\usepackage{dcolumn}% Align table columns on decimal point
\usepackage{bm}% bold math
\usepackage{epstopdf}
\usepackage{cite}
\usepackage{color}

\begin{document}

\title{Compressed sensing with sparse, structured matrices}

\author{\authorblockN{Maria Chiara Angelini }
\authorblockA{Dip. Fisica\\
Universit\`a La Sapienza\\
P.le Aldo Moro 5, 00185 Roma, Italy\\
Email: Maria.Chiara.Angelini@roma1.infn.it}
\and
\authorblockN{Federico Ricci-Tersenghi}
\authorblockA{Dip. Fisica, CNR -- IPCF, UOS Roma\\
INFN -- Roma 1, Universit\`a La Sapienza\\
P.le Aldo Moro 5, 00185 Roma, Italy\\
Email: Federico.Ricci@roma1.infn.it}
\and
\authorblockN{Yoshiyuki Kabashima }
\authorblockA{Dept. of Comput. Intell. \& Syst. Sci.\\
Tokyo Institute of Technology\\
Yokohama 226-8502, Japan\\
Email: kaba@dis.titech.ac.jp}}

\maketitle

\begin{abstract}
In the context of the compressed sensing problem, we propose a new ensemble of sparse random matrices which allow one (i) to acquire and compress a $\rho_0$-sparse signal of length $N$ in a time linear in $N$ and (ii) to perfectly recover the original signal, compressed at a rate $\alpha$, by using a message passing algorithm (Expectation Maximization Belief Propagation) that runs in a time linear in $N$. In the large $N$ limit, the scheme proposed here closely approaches the theoretical bound $\rho_0=\alpha$, and so it is both optimal and efficient (linear time complexity). More generally, we show that several ensembles of dense random matrices can be converted into ensembles of sparse random matrices, having the same thresholds, but much lower computational complexity.
\end{abstract}

\IEEEpeerreviewmaketitle

\section{Introduction}
Compressed sensing is a framework that enables an $N$-dimensional sparse signal $\bm{s}=(s_i)$ to be recovered from $M(<N)$ linear measurements of its elements, $\bm{y}=\bm{Fs}$, by exploiting the prior knowledge that $\bm{s}$ contains many zero elements \cite{intro_CS}. A simple consideration guarantees that $\ell_0$-recovery, 
\begin{eqnarray}
\hat{\bm{s}}=\mathop{\rm argmin}_{\bm{x}} || \bm{x} ||_0 \quad
\mbox{subj. to} \quad\bm{y}=\bm{F} \bm{x}, 
\label{l0}
\end{eqnarray}
where $||\bm{x}||_0$ denotes the number of non-zero elements in $\bm{x}$, is theoretically optimal in terms of minimizing the number of measurements $M$ necessary for perfectly recovering any original signal $\bm{s}$. However, carrying out $\ell_0$-recovery for a general measurement matrix $\bm{F}$ is NP-hard. To avoid such computational difficulties, an alternative approach, $\ell_1$-recovery
\begin{eqnarray}
\hat{\bm{s}}=\mathop{\rm argmin}_{\bm{x}} || \bm{x} ||_1 \quad
\mbox{subj. to} \quad \bm{y}=\bm{F} \bm{x}, 
\label{l1}
\end{eqnarray}
where $||\bm{x}||_1=\sum_{i=1}^N |x_i|$, is widely employed, as (\ref{l1}) is generally converted into a linear programming problem, and therefore, signal recovery is mathematically guaranteed in an $O(N^3)$ computational time through the use of the interior point method. Nevertheless, the $O(N^3)$ cost of computation can still be unacceptably high in many practical situations, and much effort is being put into finding more computationally feasible and accurate recovery schemes \cite{OMP,IRLS,Bregman,IHT,AMP,Krzakala}. 

Among such efforts, the recovery scheme recently proposed by Krzakala et al.~\cite{Krzakala} is worth special attention. Their scheme basically follows the Bayesian approach. Namely, the signal recovery problem is formulated as one of statistical inference from the posterior distribution, 
\begin{eqnarray}
P(\bm{x}|\bm{F},\bm{y})
=\frac{\delta(\bm{F}\bm{x}-\bm{y}) P(\bm{x})}{Z(\bm{F},\bm{y})}, 
\label{posterior}
\end{eqnarray}
where $Z(\bm{F},\bm{y})$ is a normalization factor imposing the condition $\int d\bm{x}P(\bm{x}|\bm{F},\bm{y})=1$ and a component-wise prior distribution $P(\bm{x})=\prod_{i=1}^N [(1-\rho) \delta(x_i)+ \rho \phi(x_i)]$ is assumed. $\rho$ and $\phi(\bm{x})$ represent the density of the non-zero signal elements and a Gaussian distribution, respectively. Exactly inferring $\bm{s}$ from (\ref{posterior}) is NP-hard, similarly to (\ref{l0}). However, by employing the belief propagation (BP) in conjunction with the expectation-maximization (EM) algorithm for estimating $\rho$ and the parameters of $\phi(x)$, they developed an approximation algorithm, termed EM-BP, which has better recovery performance than (\ref{l1}) with a computational cost of only $O(N^2)$. Furthermore, they showed that, by employing a peculiar type of ``seeded'' matrix $\bm{F}$, the threshold of the compression rate $\alpha=M/N$ of EM-BP, above which the original signal is typically recovered successfully, can approach very close to that of $\ell_0$-recovery, $\alpha_{s-EMBP}=\rho_0$, where $\rho_0$ is the actual signal density of $\bm{s}$ and $s-EMBP$ stands for `seeded EM-BP'. The seeded matrix is composed of blocks along the diagonal densely filled with Gaussian random variables. It is important to remember that this result is achieved for the first time with an approach different from $\ell_0$-recovery, being the threshold for $\ell_1$-recovery that is much higher than the optimal one: $\alpha_{\ell_1}>\rho_0$. The optimality of the $\ell_0$-recovery is guaranteed for EM-BP, which means that this scheme can practically achieve the theoretically optimal threshold of signal recovery with an $O(N^2)$ computational cost. 
This remarkable property was recently proved in a mathematically rigorous manner
in the case that the matrix entries satisfy certain conditions concerning their statistics~\cite{Montanari}.
However, it is still unclear whether their scheme is optimal in terms of the computational complexity; there might be a certain design of the measurement matrix $\bm{F}$ that makes it possible to further reduce the computational cost while keeping the same signal recovery threshold. 

The purpose of the present study is to explore such a possibility. For this, we focus on a class of matrices that are characterized by the following properties:
\begin{itemize}
\item {\bf sparsity:} The matrix $\bm{F}$ has only $O(1)$ non-zero elements per row and column. This implies that the measurements can be performed in a time linear in the signal length. This situation is highly preferred for the sake of practicality, given that such an operation typically needs to be done in real time, during data acquisition. 
\item {\bf integer values:} The matrix elements are not real valued, but take on small integer values. This means that an optimized code for the measurements can work with bitwise operations, thus achieving much better performance without any loss of precision. 
\item {\bf no block structure:} The block structure used in \cite{Krzakala} may not be necessary for reaching the optimal threshold. As an alternative possibility, we study a structure made of a square matrix in the upper left corner (the seed) plus a stripe along the diagonal. This structure is much more amenable for analytic computations, since it corresponds to a one-dimensional model homogeneous in space.
\end{itemize}

The use of sparse matrices for compressed sensing has already been suggested in several earlier studies \cite{Berinde,Gilbert,Akcakaya,KabashimaWadayama,Baron}. Our approach is particularly similar to that of \cite{Baron} in the sense that both sides are based on the Bayesian framework and use integer-valued sparse measurement matrices. Nevertheless, these two approaches differ considerably in the following two points. Firstly, we adopt EM-BP, which updates only a few variables per node for the signal recovery, while the recovery algorithm of \cite{Baron} involves functional updates and needs significantly more computational time than ours. Secondly, we thoroughly explored a simple design of $\bm{F}$ that achieves nearly optimal recovery performance. In contrast, the problem of the matrix design is not fully examined in \cite{Baron}. By carrying out extensive numerical experiments in conjunction with an analysis based on density evolution \cite{RichardsonUrbanke}, we show that a threshold close to the theoretical limit $\alpha=\rho_0$ can be achieved by using matrices with the above properties with an almost {\em linear} computational cost in the measurement and recovery stages. 

This paper is organized as follows. In Section \ref{sec:EM-BP}, the EM-BP algorithm is briefly explained and the results for dense matrices are summarized. The algorithm is applied to homogeneously sparse matrices in Section \ref{sec:sparse} and to structured sparse block matrices in Section \ref{sec:block}. A new type of ``striped'' sparse matrix without blocks is introduced in Section \ref{sec:stripe}. The last Section summarizes our work, focusing on its importance for practical use, and touches on future issues. 

\section{Expectation Maximization Belief Propagation\label{sec:EM-BP}}
The new algorithm based on BP in conjunction with the EM proposed in Ref. \cite{Krzakala} starts from Eq.~(\ref{posterior}). 
A similar idea was also proposed in Ref. \cite{Schniter}.
In order to solve it with BP, $O(MN)$ messages for the probability distributions of the variables $x_i$ are constructed in the following way:
\begin{align*}
m_{\mu\rightarrow i}(x_i)&=\frac{1}{Z^{\mu\rightarrow i}}\int\prod_{j\neq i}dx_j m_{j\rightarrow \mu}(x_i) \delta\Big(y_{\mu}-\sum_k F_{\mu k}x_k\Big)\\
m_{i\rightarrow\mu}(x_i)&=\frac{1}{Z^{i\rightarrow\mu}}\Big[(1-\rho) \delta(x_i)+ \rho \phi(x_i)\Big] \prod_{\gamma\neq \mu} m_{\gamma\rightarrow i}(x_i)
\end{align*}
where $Z^{i\rightarrow\mu}$ and $Z^{\mu\rightarrow i}$ are normalization factors. This EM-BP equations are very complicated because the messages are distribution functions. In order to make them simpler, the messages can be approximated by assuming that they are Gaussian, thus obtaining the equations for the mean $a_{i\rightarrow\mu}$ and the variance $v_{i\rightarrow\mu}$ of $m_{i\rightarrow \mu}(x_i)$. This approximation was introduced for sparse matrices in Refs.~\cite{Kabashima2003,Guo,MontanariTse,Rangan}, and it becomes asymptotically exact if $\bm{F}$ is dense. 
In fact, it is derived from an expansion in small $F_{\mu i}$, and in the dense case $F_{\mu i}=O(1/\sqrt{N})$. Supposing that the elements of the original signal follow a Bernoulli-Gaussian distribution with parameters $\rho_0$, $\overline{x}_0$ and $\sigma_0$, the update rules for the messages are the following:
\begin{align}
a_{i\rightarrow\mu}&=f_a\left(\sum_{\gamma\neq\mu}A_{\gamma\rightarrow i},\sum_{\gamma\neq\mu}B_{\gamma\rightarrow i}\right) \nonumber \\ \nonumber
a_{i}&=f_a\left(\sum_{\gamma}A_{\gamma\rightarrow i},\sum_{\gamma}B_{\gamma\rightarrow i}\right)\\ \nonumber
v_{i\rightarrow\mu}&=f_c\left(\sum_{\gamma\neq\mu}A_{\gamma\rightarrow i},\sum_{\gamma\neq\mu}B_{\gamma\rightarrow i}\right) \\ \nonumber
v_{i}&=f_c\left(\sum_{\gamma}A_{\gamma\rightarrow i},\sum_{\gamma}B_{\gamma\rightarrow i}\right)\\ \nonumber
A_{\mu\rightarrow i}&=\frac{F_{\mu i}^2}{\sum_{j\neq i}F_{\mu j}^2v_{j\rightarrow\mu}}\\ 
B_{\mu\rightarrow i}&=\frac{F_{\mu i}\left(y_{\mu}-\sum_{j\neq i}F_{\mu j}a_{j\rightarrow\mu} \right)}{\sum_{j\neq i}F_{\mu j}^2v_{j\rightarrow\mu}} 
\label{Eq:updates}
\end{align}
where $f_a$ and $f_c$ are some analytical functions depending on the parameters $\rho$, $\overline{x}$ and $\sigma$. For details, see Ref.~\cite{Krzakala}.

In general, the original $\rho_0$, $\overline{x}_0$ and $\sigma_0$ are not known, but one can use EM to derive the update rules for them, using the property that the partition function
\begin{equation*}
Z(\rho,\overline{x},\sigma)=\int d\bm{x} P(\bm{x})\delta(\bm{y}-\bm{Fx})
\end{equation*}
is the likelihood of the parameters $(\rho,\overline{x},\sigma)$ and is maximized by the true parameters $\rho_0$, $\overline{x}_0$, and $\sigma_0$. Thus, after the update of all the messages, the inferred parameters of the original distribution are updated following these rules:
\begin{align*}
\overline{x}&\leftarrow\frac{1}{\rho N}\sum_i a_i\;, \qquad
\sigma^2\leftarrow\frac{1}{\rho N}\sum_i (v_i+a_i^2)-\overline{x}^2\;,\\
\rho&\leftarrow\frac{\sum_i \frac{1/\sigma^2+U_i}{V_i+\overline{x}/\sigma^2}a_i}{\sum_i\left[1-\rho+\frac{\rho}{\sigma(1/\sigma^2+U_i)^{\frac{1}{2}}}e^{\frac{(V_i+\overline{x}/\sigma^2)^2}{2(1/\sigma^2+U_i)}-\frac{\overline{x}^2}{2\sigma^2}}\right]^{-1}}\;,
\end{align*}
with $U_i=\sum_{\gamma}A_{\gamma\rightarrow i}$ and $V_i=\sum_{\gamma}B_{\gamma\rightarrow i}$.
If the algorithm converges to the correct solution, $a_i=s_i$ and $v_i=0$.

To reduce the number of messages from $O(NM)$ to $O(N)$, one can see that in the large $N$ limit, the messages $a_{i\rightarrow\mu}$ and $v_{i\rightarrow\mu}$ are nearly independent of $\mu$. Thus, we can derive the equations involving only a variable per each measurement node and a variable per signal node, if we are careful to keep the correcting Onsager reaction term as in the TAP equations of statistical physics \cite{TAP}. This method was introduced in the context of compressed sensing in Ref. \cite{AMP} and is called approximated message passing (AMP). 

In general, the correct distribution of the original signal is unknown. However, in Ref. \cite{Krzakala}, it is demonstrated that if $\alpha>\rho_0$, the most probable configuration of $\bm{x}$ with respect to $P(\bm{x})=\prod_{i=1}^N [(1-\rho) \delta(x_i)+ \rho \phi(x_i)]$ with $\rho<1$, restricted to the subspace $\bm{y}=\bm{Fx}$, is the original signal $\bm{s}$, even if the signal is not distributed according to $P(\bm{x})$. So our choice of a Gaussian distribution for $\phi(x)$ should be perfectly fine even if the original signal has a different distribution.

The free entropy $\Phi(D)$ at a fixed mean square error $D=(1/N)\sum_{i=1}^N(x_i-s_i)^2$ can be computed if a dense matrix is used. For $\alpha>\rho_0$, the global maximum of the function $\Phi(D)$ is at $D=0$, that corresponds to the correct solution. However, below a certain threshold $\alpha<\alpha_{BP}$ that depends on the distribution $P(\bm{s})$, the free entropy develops a secondary, local maximum at $D\neq0$. As a consequence, the EM-BP algorithm can not converge to the correct solution for $\rho_0<\alpha<\alpha_{BP}$, because a dynamical transition occurs. Nonetheless, the threshold $\alpha_{BP}$ is lower than $\alpha_{\ell_1}$. 

\section{EM-BP with a sparse matrix\label{sec:sparse}}

\begin{figure}[t]
\centering
\includegraphics[width=0.8\columnwidth]{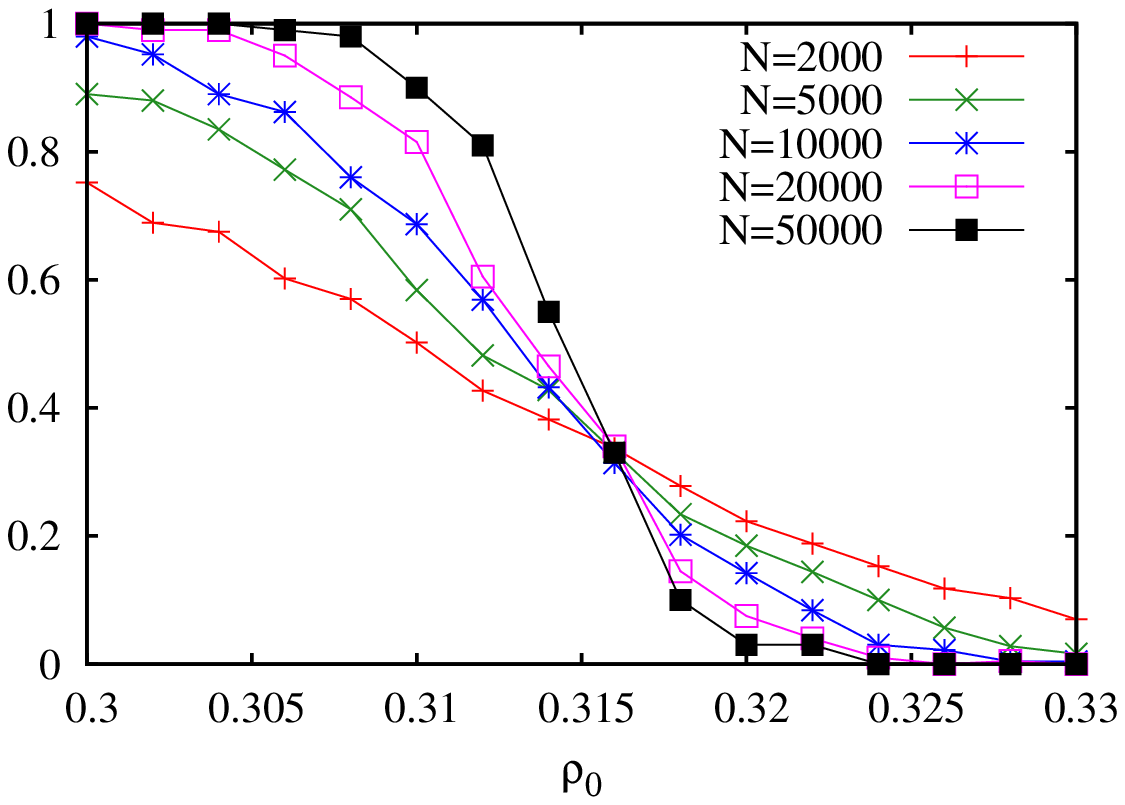}
\includegraphics[width=0.8\columnwidth]{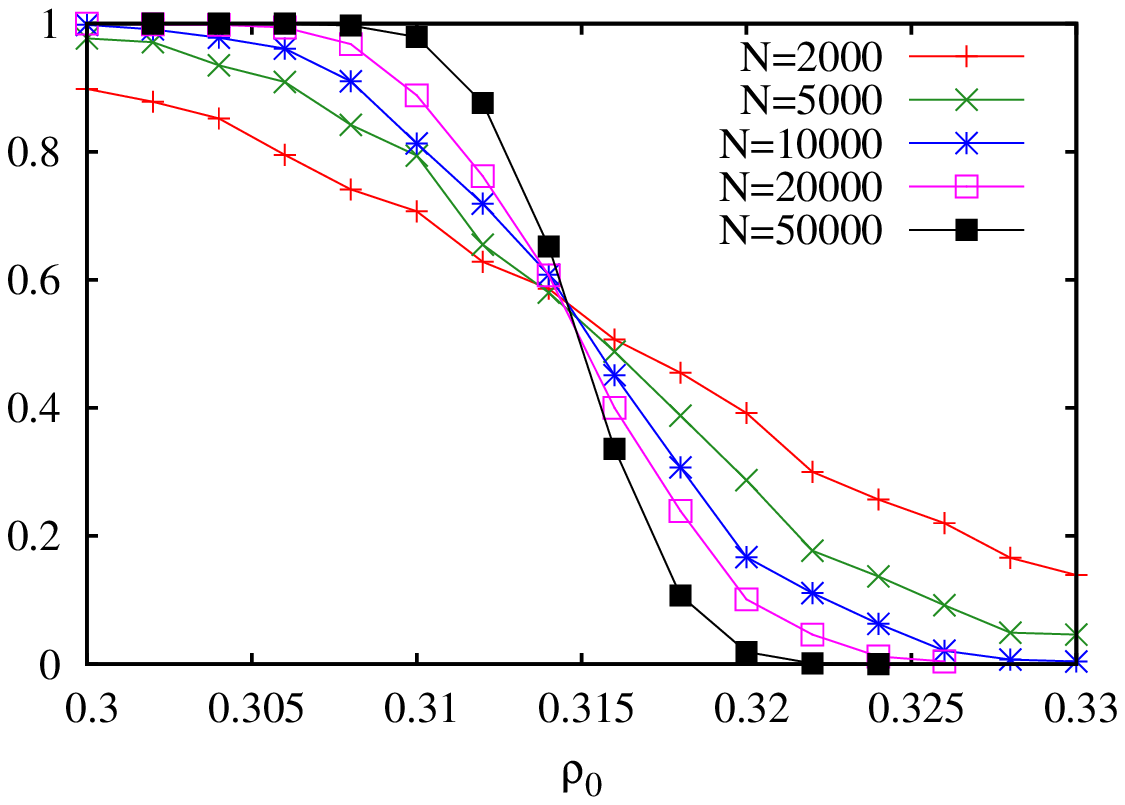}
\caption{\label{sparse}Top: Probability of perfect recovery versus the signal sparsity $\rho_0$ using sparse matrices with $\alpha=0.5$ and $K=20$. The threshold is the same as with dense matrices. Bottom: Probability of perfect recovery computed with density evolution has the same threshold. Here, $N$ is the population size.}
\end{figure}
 
First of all, we want to verify if the use of a sparse matrix can reach the same results as the use of a dense one. For a sparse random matrix, the AMP equations can not be used; thus, we can use the update rules in Eq.~(\ref{Eq:updates}) for inferring the original signal $\bm{s}$. In particular, we choose the matrix $\bm{F}$ to have only $K=O(1)$ elements different from zero in each row and $H=\alpha K=O(1)$ elements in each column, extracting them from the distribution, 
\begin{eqnarray}
P(F_{\mu i})=\frac{1}{2}\delta(F_{\mu i}-J)+\frac{1}{2}\delta(F_{\mu i}+J)
\label{Eq:PF}
\end{eqnarray}
with $J=1$. The use of the messages $a_{i\rightarrow\mu}$ and $v_{i\rightarrow\mu}$ instead of the AMP equations does not involve an extra cost in memory, because the number of the messages is $O(N)$ from the sparsity of the matrix. In principle, the messages $m_{i\rightarrow\mu}(x)$ are not Gaussian if the matrix is sparse, so the use of only the two parameters $a_{i\rightarrow\mu}$ and $v_{i\rightarrow\mu}$ is not exact. However, the convolution of $K$ messages (with $K=20$ in a typical matrix we use) is not far from a Gaussian, and indeed, we can verify a posteriori that this approximation is valid, because it gives good results.

In all our numerical simulations, we use a Bernoulli-Gaussian distributed signal and a compression rate $\alpha=0.5$.

Figure~\ref{sparse} (top) shows the probability of perfect recovery as a function of the sparsity of the signal $\rho_0$ for different sizes, by applying the EM-BP algorithm using a sparse matrix with $K=20$. The threshold for perfect recovery in the thermodynamic limit ($N\to\infty$) is $\rho_{BP}\simeq 0.315$, which is the same as the one obtained in Ref.~\cite{Krzakala} with a dense matrix. We can not analytically compute the free entropy $\Phi(D)$ as in \cite{Krzakala}, because we use sparse matrices and cannot use methods such as the saddle point one. 
However, we performed a numerical density evolution analysis, as shown in Fig.~\ref{sparse} (bottom), and found that the threshold is almost the same as the one computed with the matrices $\bm{F}$.

\begin{figure}[t]
\includegraphics[width=0.9\columnwidth]{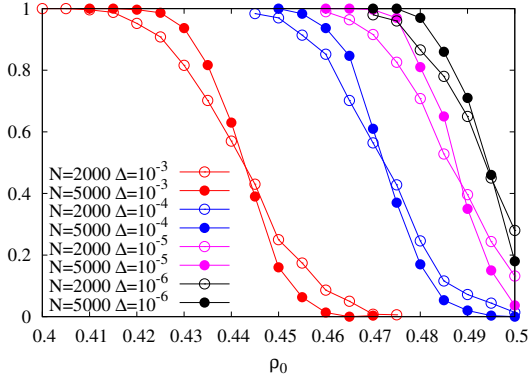}
\caption{\label{stability}Stability check of the solution determined by the EM-BP message passing algorithm. Starting the recovery process with a sparse matrix from an initial condition differing less than $\Delta$ from the correct solution, the latter is recovered as long as $\rho_0 < \alpha_{stab}(\Delta)$. In the limit $\Delta\to 0$, the stability limit $\alpha_{stab}(\Delta)$ tends to the theoretical bound $\alpha$ (which is 0.5).}
\end{figure}

Next, we will verify that the correct solution is always the global maximum of $\Phi(D)$ and it is locally stable up to $\alpha \simeq \rho_0$ when using EM-BP with a sparse matrix. Since we can not analytically compute the free entropy, we must resort to a numerical method. We start EM-BP with an initial condition very close to the correct solution: $a_{i\rightarrow\mu}^0=s_i+\delta_{i\rightarrow\mu}$, with $\delta_{i\rightarrow\mu}$ a random number uniformly distributed in $[-\Delta,\Delta]$. In this way, we have verified (see Fig.~\ref{stability}) that if $\Delta$ is sufficiently small, the correct solution can be found up to $\alpha \simeq \rho_0$, as in the case of a dense matrix.

For the algorithms based on the $\ell_1$ minimization, it is known that the threshold with a sparse matrix is lower than that with a dense one. However, these algorithms are not optimal, because the correct solution disappears below the threshold $\alpha_{\ell_1}$. In this sense, the EM-BP algorithm is optimal, because the global maximum of the free entropy is always on the correct solution. Thus, one can expect that, if the rank of the sparse matrix is the same as that of the dense one, a similar threshold can be reached (as we have demonstrated numerically).

In summary, we can say that the EM-BP algorithm of Ref. \cite{Krzakala} seems to reach the same threshold $\alpha_{BP}$, either using a dense Gaussian matrix or a sparse binary one
if the numbers of non-zero elements per row/column are $O(1)$ but sufficiently large.
However, the use of a sparse matrix is computationally much faster than the use of a dense one. Moreover, the use of binary elements, instead of Gaussian real values, allows for better code optimization and eventually for hard-wired encoding of the compression process.

\section{Block-structured sparse matrices\label{sec:block}}

\begin{figure}[t]
\includegraphics[width=\columnwidth]{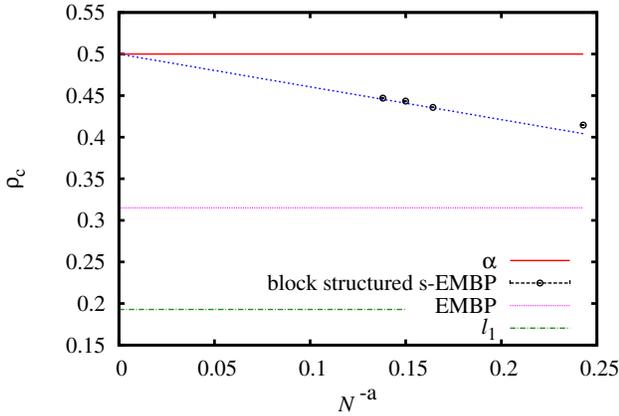}
\caption{\label{structured}$\rho_c(N)$ with a sparse, structured matrix with blocks, for different sizes $N$ and values of $L$ (see text). The thresholds for the $\ell_1$-recovery and for EM-BP without any structure are also drawn. For comparison with the data in Fig.~\ref{unid}, we used the exponent $a \simeq 0.18$ best fitting those data.}
\end{figure}

To avoid the secondary maximum of the free entropy, in Ref. \cite{Krzakala}, the authors use a structured block matrix that helps to nucleate the correct solution. The idea is that the correct solution is found for the first variables, and then it propagates to the whole signal. This idea is similar to the so-called spatial coupling that is very useful for solving many different problems \cite{spat-coup}. With this trick, the authors of Ref.~\cite{Krzakala} reach perfect recovery for almost any $\alpha > \rho_0$ in the large $N$ limit while Ref.~\cite{Kudekar} reports that the gain is quite small when different recovery algorithms are used. Here, we try to use a matrix with the same block structure, but sparsely filled. We divide the $N$ variables into $L$ groups of size $N/L$ and $M$ measurements into $L$ groups of size $M_p=\alpha_pN/L$ in such a way that $M=\sum_{p=1}^L M_p=\alpha N$ and $1/L\sum_{p=1}^L\alpha_p=\alpha$.

In this way, the matrix ${\bm F}$ is divided into $L^2$ blocks, labeled with indices $(p,q)$. Each block is a sparse binary matrix with $k$ elements different from zero for each row and $h_p=\alpha_p k$ elements for each column, distributed according to Eq.~(\ref{Eq:PF}), with $J=J_{p,q}$. As in Ref. \cite{Krzakala}, we choose $J_{p,p-1}=J_1$, $J_{p,p}=1$, $J_{p,p+1}=J_2$, and $J_{p,q}=0$ otherwise. The important ingredient to nucleate the correct solution is that in the first block $\alpha_1=(M_1/N)L>\alpha_{BP}$ holds. For simplicity, we can choose $\alpha_1=1$ and $\alpha_p=(L\alpha-1)/(L-1)$ for $p\neq1$. 
The recovery strongly depends on the parameters $J_1$ and $J_2$, and the best results for $\alpha=0.5$ are obtained around $J_1=4$ and $J_2=1$. Moreover, we used these two values in the experiments described below because we wanted to work with matrices with elements having small integer values.

Similarly to the dense case, the use of a sparse structured matrix with blocks allows to overcome the dynamical transition at $\alpha_{BP}$ and to nucleate the correct solution until $\alpha$ is very close to $\rho_0$. Figure~\ref{structured} shows the mean critical threshold $\rho_c(N)$ for different signal lengths at a fixed compression rate $\alpha=0.5$. The $x$ axis uses the same scaling variable as in Fig.~\ref{unid}, and the best parameter $a$ obtained from the fit of data in Fig.~\ref{unid} also interpolates the data in Fig.~\ref{structured} quite well. In the thermodynamic limit, $\rho_c$ extrapolates to a value compatible with the optimal one, $\alpha$, and it is certainly much higher than the thresholds for $\ell_1$-recovery and for EM-BP without any structure. We have also done a density evolution analysis that confirms this result.  

For each value of $N$ and $L$, the mean critical threshold $\rho_c$ is computed as follows. We randomly generate a block structured matrix ${\bm F}$ with the given $N$ and $L$. We start with a sufficiently sparse original signal $\bm{s}$, which has been recovered by the algorithm; we then add non-zero entries to the signal and check whether the new signal can be recovered by the algorithm; we go on adding non-zero elements to the signal until a failure in a perfect recovery occurs. The previous to the last value for $\rho_0$ is the critical threshold for the matrix ${\bm F}$. The mean critical threshold is obtained by averaging over many different random matrices and signals, with the same values of $N$ and $L$. The number of such random extractions goes from $10^3$ for the largest $N$ value up to $10^4$ for the smallest $N$ value.

The values of $(N,L,k)$ used for the simulations shown in Fig.~\ref{structured} are the following: $(2250,10,9)$, $(19000,20,19)$, $(31200,40,39)$, and $(49000,50,49)$. We need to increase both $N$ and $L$ if we want to obtain good results in the thermodynamic limit. However, if we change $L$, we must change $k$ too. Indeed, in order to have the same number of elements per row and column for each of the $L^2$ blocks, we must satisfy the conditions: $(N/L)/M_p=k/h_p$ with $k$ and $h_p$ integer valued. Here, we have used the smaller possible value for $k$, that is $k=L-1$. The fact that it is impossible to keep $k$ constant while increasing $L$ implies that these kinds of block-structured matrices always become dense in the thermodynamic limit. This is a limitation of the block structure that we want to eliminate with the matrix proposed in the following Section.

\section{And without blocks?\label{sec:stripe}}
The matrix proposed in Ref. \cite{Krzakala} is not the only one that allows the optimal threshold to be reached. Reference \cite{Krzakala2} analyzes the use of other good dense, block-structured matrices. However, the block-structure is not so simple to handle if one wants to do analytical calculations in the continuum limit. Moreover, in making these block-structured matrices sparse, one has to be careful to find the right values of $L, M, N, M_p, \alpha_p, k$. For these reasons, we want to know if the block structure is crucial, and, if not, we want to eliminate it.

\begin{figure}[t]
\includegraphics[width=\columnwidth]{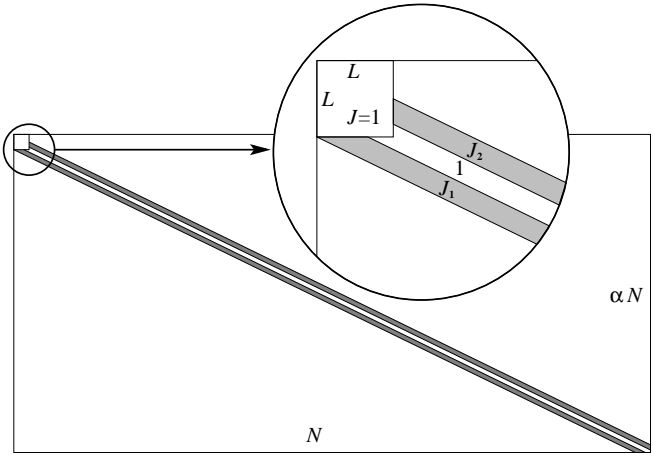}
\caption{\label{matrixunid}A nearly one-dimensional sparse matrix with a square block of size $L \times L$ at the top left and non-zero elements in the stripes around the diagonal can achieve compression and perfect recovery close to the theoretical bound in linear time.}
\end{figure}

We tried a different structured sparse matrix (see Fig.~\ref{matrixunid}), that we called a striped matrix. It has one sparse square block of size $L$ on the top left of the matrix with $K=O(1)$ elements for each row and column extracted from (\ref{Eq:PF}) with $J=1$. This arrangement is fundamental for nucleating the correct solution. Apart from this first block, the residual compression rate is $\alpha'=\frac{M-L}{N-L}$. Then, we construct a one-dimensional structure around the diagonal of the remaining matrix. For each column $c>L$, we randomly place $2K\alpha'$ non-zero elements, again extracted from (\ref{Eq:PF}), in the interval of the width $2L\alpha'$ around the diagonal. One element with $J=1$ is always placed on the diagonal (actually on the position closest to the diagonal). For the remaining elements, we use the following rules. If the element is at a distance $d\leq L\alpha'/3$ from the diagonal, we use $J= 1$. Otherwise, if its distance is $d>L\alpha'/3$, we use $J=J_1$ below the diagonal and $J=J_2$ above the diagonal. In this way, the number of elements per column is constant, while the number of elements per row is a truncated Poisson random variable with mean $2K$: indeed, there are no empty rows, thanks to the rule of placing the first element of each column closest to the diagonal. When constructing the matrix, we apply exactly the same rule to each column, but in the last $L$ columns it may happen that a non-zero element has a row index larger than $M$: these elements are then moved below the first square matrix by changing the row and column indices as follows: $r \leftarrow r - (M-L)$ and $c \leftarrow c - (N-L)$.

\begin{figure}[t]
\includegraphics[width=\columnwidth]{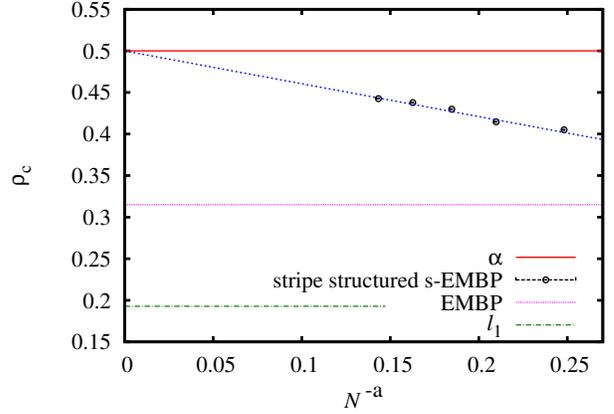}
\caption{\label{unid}$\rho_c(N)$ with a sparse, striped matrix, as described in Section \ref{sec:stripe} for different sizes (from $N=2000$ to $N=40000$). The thresholds for $\ell_1$-recovery and for EM-BP without structure are drawn for comparison. The best fitting parameter is $a \simeq 0.18$, and it leads to an extrapolation of $\rho_c(N)$ in the thermodynamic limit compatible with $\alpha$.}
\end{figure}

In this way, we have some kind of continuous one-dimensional version of the block-structured matrix discussed in the previous Section. Within this striped matrix ensemble, the thermodynamic limit at a fixed matrix sparsity can be calculated without any problem, by sending $N,L \to \infty$ at a fixed $L/N$ and fixed $K=O(1)$. In Fig.~\ref{unid}, we show the mean critical threshold reached by using striped matrices with a fixed ratio $L/N=1/50$ (the same used in the plot of Fig.~\ref{matrixunid}) and different signal lengths. Perfect decoding up to $\rho_c$ is again achieved by using the EM-BP algorithm. We extrapolated the $\rho_c(N)$ data to the thermodynamic limit by assuming the following behavior in the large $N$ limit:
\begin{equation}
\rho_c(N) = \rho_c(\infty) - b N^{-a}
\end{equation}
The data in Fig.~\ref{unid} are plotted with the best fitting parameter $a \simeq 0.18$, and the extrapolated value $\rho_c(\infty)$ is perfectly compatible with the theoretical bound $\alpha$.

Hence, we can conclude that the important ingredient to reach optimality is not the block structure, but the nearly one-dimensional structure, associated with the initial block with $\alpha_1 > \alpha_{BP}$ to nucleate the correct solution.

It is worth noticing that the corresponding statistical mechanics model for these striped random matrices is a one-dimensional disordered model with an interaction range growing with the signal length, as in a Kac construction. Models of this kind are analytically solvable and have shown very interesting results \cite{SilvioAndrea}.

\begin{figure}[t]
\includegraphics[width=0.9\columnwidth]{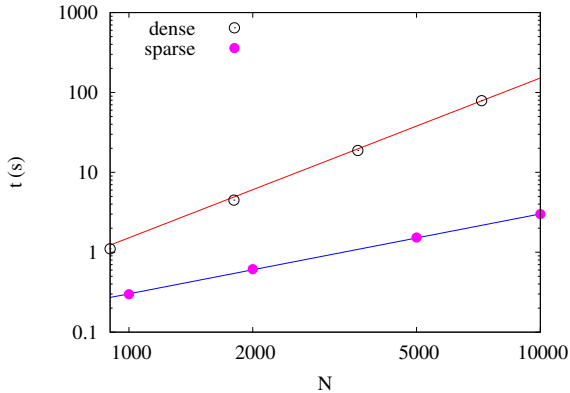}
\caption{\label{Fig:time}Actual time (in seconds) for recovery of a signal with dense and sparse matrices for different data lengths $N$. The data are fitted respectively by a quadratic and a linear function.}
\end{figure}

The use of our striped sparse matrices allows for a great reduction in computational complexity. Indeed, the measurement and recovery times grow linearly with the size of the signal if sparse matrices are used, while they grow quadratically if dense matrices are used. Figure \ref{Fig:time} shows the measurement and recovery times of a signal for different signal lengths $N$. For this test, we used dense block-structured matrices and sparse striped matrices. The number of EM-BP iterations to reach the solution is roughly constant for different $N$. A quadratic fit for the dense case and a linear fit for the sparse one perfectly interpolate the data.

\section{Conclusions and future development\label{sec:conclusions}}
We introduced an ensemble of sparse random matrices ${\bm F}$ that, thanks to their particular structure (see Figure \ref{matrixunid}), allow us to perform the following operations in linear time:\\ 
(i) measurement of a $\rho_0$-sparse vector ${\bm s}$ of length $N$ by using a linear transformation, ${\bm y} = \bm{Fs}$, to a vector ${\bm y}$ of length $\alpha N$ \\
(ii) perfect recovery of the original vector ${\bm s}$ by using a message passing algorithm (Expectation Maximization Belief Propagation) for almost any parameter satisfying the theoretical bound $\rho_0 < \alpha$.

These striped sparse matrices that have such good performance because there is a `seeding' sparse square matrix in the upper left corner that nucleates a seed for the right solution and the one-dimensional structure along the diagonal propagates the initial seed to the complete right solution. Both seeding and a one-dimensional structure have been used in the past \cite{spat-coup,Krzakala}, but in our new ensemble, the matrices are sparse, and this permits us to perform all the operations in a time linear in the signal length.

We also checked that sparse matrices perform as well as dense ones in the case of block-structured matrices and for matrices with no structure at all.

Apart from the compressed sensing case, several other applications require a sparse matrix or equivalently linear time complexity \cite{Gilbert}

In data streaming computing, one is typically interested in doing very quick measurements in constant time. For example, if the task is to measure the number of packets $s_i$ with destination $i$ passing through a network router, it is not possible to keep a vector $\bm{s}$ because it is generally too long. Instead, a much shorter sketch of it, $\bm{y}=\bm{Fs}$, is measured in such  a way that a very sparse vector $\bm{s}$ can be recovered from $\bm{y}$. The matrix $\bm{F}$ must be sparse in order to be able to update the sketch $\bm{y}$ in a constant time for each new packet passing through the router.

Another interesting application is the problem of group testing, where a very sparse vector $\bm{s}\in\{0,1\}^N$ is given and one is interested in performing the fewest linear measurements, $\bm{y}=\bm{Fs}$, that allow for detection of the defective elements ($s_i=1$). In this case, the experimental constraints require a sparse matrix $\bm{F}$: only if the tested compound $y_\mu = \sum_i F_{\mu i} s_i$ is made of a very few elements of $\bm{s}$, the linear response holds and non-linear effects can be ignored.

However, in the more general case, one does not directly observe the sparse signal $\bm{s}$ but rather a linear transformation of it, $\bm{x}=\bm{Ds}$, made with a dictionary matrix $\bm{D}$ (which is typically a Fourier or wavelet transformation, and thus is a dense matrix). In this more difficult case, one would like to design a sparse measurement matrix $\bm{A}$ such that the measurement/compression operation, $\bm{y}=\bm{Ax}$, is fast, and the resulting observed data $\bm{y}$ is short, thanks to the sparseness of $\bm{s}$. The conflicting requirement is to have a fast recovery scheme, because, now, to recover the original signal one should solve $\hat{\bm{s}}= {\rm argmin}\,||\bm{s}||_0$ subject to $\bm{y}=(\bm{AD})\bm{s}$, where $\bm{AD}$ is typically dense (e.g.\ in case of Fourier and wavelet transformations). So a very interesting future development of the present approach is to extend it to this more complex case.

\section*{Acknowledgments}
FR-T is grateful for his useful discussions with F.~Krzakala, M.~M\'ezard and L.~Zdeborova and financial support from the Italian Research Minister through the FIRB Project No. RBFR086NN1 on ``Inference and Optimization in Complex Systems: From the Thermodynamics of Spin Glasses to Message Passing Algorithms''.

YK is supported by grants from the Japan Society for the Promotion of Science (KAKENHI No. 22300003) and the Mitsubishi foundation.

\IEEEtriggeratref{16}

\end{document}